# A Simple Mutual Information based Registration Method for Thermal-Optical Image Pairs applied on a Novel Dataset

Suranjan Goswami
CVBL
IIITA
Prayagraj, India
https://orcid.org/0000-0002-5499-8748

Satish Kumar Singh
CVBL
IIITA
Prayagraj, India
sk.singh@iiita.ac.in

*Abstract*— **While thermal optical registered datasets are becoming widely available, most of these works are based on image pairs which are pre-registered. However, thermal imagers where these images are registered by default are quite expensive. We present in this work, a thermal image registration technique which is computationally lightweight, and can be employed regardless of the resolution of the images captured. We use 2 different thermal imagers to create a completely new database and introduce it as a part of this work as well. The images captured are based on 5 different classes and encompass subjects like the Prayagraj Kumbh Mela, one of the largest public fairs in the world, captured over a period of 2 years.**

## I. Introduction

Thermal Images can be widely grouped into 2 different categories, namely Near Infrared (NIR) images and the Thermal Infrared (TIR) images. The difference between the 2 is that NIR images incorporate features from their optical counterparts in the images, while TIR images are based purely on the passive radiation of the bodies. Thus, NIR images, by virtue of their shared optical image properties, are much easier to work with. We present in this work, a new registration method based on the Mutual Information (MI), which works on TIR images.

We also introduce a new unique database [1] as a part of this work, which is based on 5 different classes of data, human, modern infrastructure, animal, greenery and crowd. The images were captured via 2 different thermal imagers, FLIR E40 and Sonel KT400. The motivation for this work is, despite the fact that there are several works on thermal images like [2-4], all of them use a database [5] which is a single domain database, comprising purely of pedestrian images captured from a video. This means that all of these works are using a dataset comprised of images which are very similar to each other. The low variance and high similarity of such images provide an inherent bias which might not be suitable for works based on generalized domains. We found one other instance of a large scale thermal image dataset, from the Military Sensing Information Analysis Center (SENSIAC) [6], which was used for thermal image colorization in [7]. The data that they provide is in the form of military specific data obtained from soldiers, military vehicles and civilian vehicles. Our work, however, is geared towards using data obtained wholly from civilian scenarios, so that it may be applied directly to data obtained in daily life. Also, military data is often obtained via very high-grade state of the art cameras, which are not feasible for general use. Finally, the dataset is not publicly available, and needs a fee for its usage.

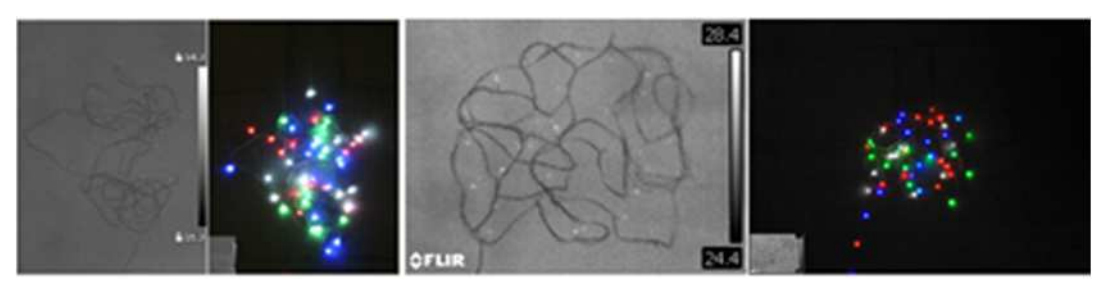

Fig. 1. A comparison of the sizes of the thermal image vs the optical image for a scene captured via 2 different thermal imagers. Figure (a) is captured via Sonel KT400 and figure (b) is the image captured via FLIR E40. The thermal images for each are inset on the bottom left of the optical images for a visual comparison.

All of the above works are based on the capture of the images, and none of them actually talk about the registration module needed. While Yan et al. [8] does touch on the aspect of thermal image registration, we found that when we used real world images, we could simplify the algorithm. Not only that, their work does not focus on the aspect of image scaling either, where they assumed that the image pairs would be of the same resolution. Additionally, if there is no problem of scaling present in a thermal imager, it reduces to just a centre-shift problem, which is simpler than the model we are proposing here. While this is similar to multi modal registration methods, most of such works like [9] focus on additional factors like corner detectors or use of localized voxels. The method we are proposing is simpler in terms of computational complexity, while showing similar results.

## II. Proposed Method

The registration method is created to find the optical counterpart of a given thermal image. Besides problems like random centre shift, images obtained via the optical sensor and the thermal sensors might not have same information because images captured in low light or similar temperature profiled regions might yield different images. As such, standard optical matching techniques like Structural Similarity Index Measure (SSIM), which work by matching how similar 2 images are, would fail. Illumination based

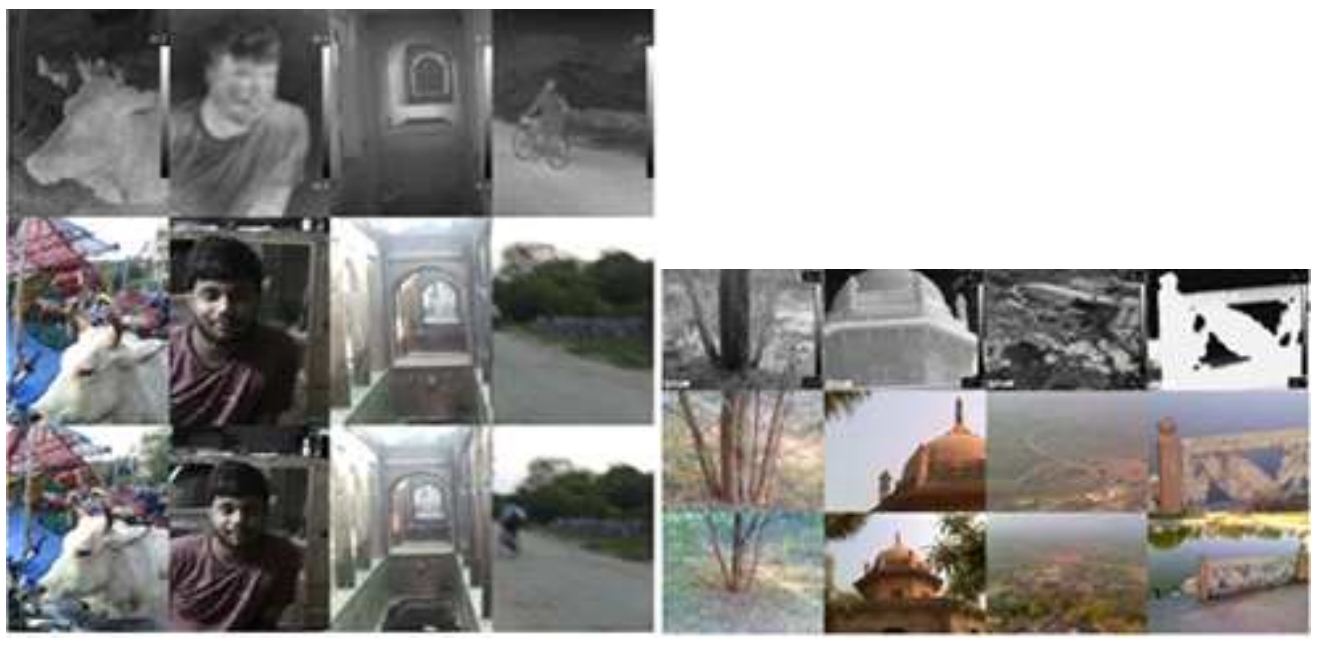

Fig. 4. Images where the Registration algorithm is unable to provide a perfect match of thermal-optical pairs in case of images captured via the Sonel thermal imager for (a) – (d) and those captured by the FLIR thermal imager for (e) - (h)

,

*Algorithm 2* and *Algorithm 3* for registration of thermal-optical pairs. RA stands for Registration Algorithm.

similarity scores like Histogram Matching would, on the other hand, fail, because the illumination levels available via the 2 sensors are completely different from each other. However, we reasoned that the internal entropy of an image should be same regardless of the sensor used to capture the image since it represents the amount of disorder or information content in an image. Thus, we opted for Mutual information (MI) as a matching score for the registration metric in our method. While there exists a prior work [10] which uses MI for thermal image registration, their work corrected the illumination in the input images before the registration. We, however reasoned that this would not yield any extra benefits since MI works with the relative distribution of the illumination, rather than with the absolute values. . Of course, our method would still fail in case of complete darkness due to the complete absence of any optical counterparts to the thermal images, but that is an inherent problem with any registration method.

We use the FLIR E40 imager with a 3.1 Megapixel optical sensor producing images at a resolution of 1536x2048 pixels and the Sonel KT400 imager with a 5 Megapixel optical sensor producing images at a resolution of 2592x1944 pixels. Their respective thermal image sizes are 240x320 and 384x288. This means that the images are in landscape for the FLIR imager and in portrait mode for the Sonel imager. We used the concept of Homography to solve this problem, where, in order to calculate the corresponding region that maps each $x_i$ (data in the first domain) to its corresponding $X_i$ (data in the second domain), we need to simply compute the 3×3 homography matrix [11].

For obtaining the 4 corresponding points in 2 images from 2 different domains, we took images where we could find out direct pixel values from images obtained across different domains for the same scene. This is shown in Fig. 1, where we use LED lighting strips in order to create basic images. In that figure, (a) is the image obtained via the Sonel thermal imager, and (b) is the one obtained via the FLIR thermal imager. Both optical images have an inset of the original thermal image at the bottom left to give an idea of the difference in scales between the captured thermal and optical images. . The rescaling factor is 18% in case of the images captured via the Sonel thermal imager and 36.5% in case of the images captured via the FLIR thermal imager, uniformly across both axes.

We outline the algorithm (*Algorithm 1*) proposed for our Registration method as below and we call it *Registration Algorithm 1*. In it, N stands for the total size of the flattened rescaled optical grayscale image. K stands for the size of the thermal image, which is 384x288 = 110,592 for images captured via the Sonel imager and 240x320 = 76,800 for images captured via the FLIR imager. Once the region is calculated, the corresponding 2D index has to be calculated for the point on the resized optical image for obtaining the registered image, followed by reshaping it to the original 2D shape from 1D array.

Algorithm 1

**Input**: Thermal-Optical image pair
**Output**: Cropped region of registered Optical image
*for each image pair in database:*
1. *flatten the thermal image as thermal*
2. *flatten the grayscale image as grayscale*
3. *temp = 0*
4. *for i in 0 to N-K:*
   4.1. *if (temp < MI (thermal, grayscale[i to i+K]):*
      4.1.1. *x =i*
      4.1.2. *temp = MI (thermal, grayscale[i to i+K])*
5. *save image corresponding to x from grayscale image in 2D format*

In the above algorithm, *temp* represents the score and *MI* represents the Mutual Information function, given 2 distributions.

In mathematical terms, the registration process becomes an optimization problem based on Shannon's entropy, which can be expressed as below. Given a matrix of the thermal image *A* with shape *(a, b)* and an optical image matrix *X* with shape *(x, y)*, we can define a patch $p_i$ where

$$p_i = (x_i : x_i + a, y_i : y_i + b) \tag{1}$$

as the *(a , b)* patch with the highest Mutual Information (MI) in the optical image and $(x_i, y_i)$ being the starting index of the patch inside the image, where

$$0 \leq a_i \leq x - a, \quad 0 \leq b_i \leq y - b \tag{2}$$

since the values of *a* and *b* has to be bound between 0 and the last indices subtracted from the size of the image being considered for registration. We calculate the region with the highest Mutual Information in the optical image *X* based on the Mutual Information (MI) formula following *Algorithm 1*, where MI is defined as:

$$MI = H(A) + H(X) - H(A, X)$$

$$= \sum_{a \in A} \sum_{x \in X} p_{AX}(a, x) \log \frac{p_{AX}(a,x)}{p_A(a) p_X(x)} \tag{3}$$

where *H(X)* represents the entropy of system *X*. However, the calculation of MI is a very computationally expensive work.

As such, we simplified the method by reducing the number of checks that need to be performed as a part of the search problem. Instead of searching the full optical image for the most optimal region corresponding to the thermal image, we change the range for the check to half of the difference between the 2 images. Thus, the range reduces from $(0 : x - a , 0 : y - b)$ to $( (x-a)/2 - (x-a)/4 : (x-a)/2 + (x-a)/4 , (y-b)/2 - (y-b)/4: (y-b)/2 + (y-b)/4 )$.

We make an assumption here that the maximum amount of shift in pixel would not be more than half the total difference in sizes between the 2 images. This is also shown to be valid experimentally for both imagers when we take an image at a very close range and check the parallax error present between the thermal and the resized optical images. We show this as Fig. 5.

However, while we can reduce the time complexity by about 50%, we found that there were a lot of redundant checks being performed as a part of the checking procedure because after the maximal region is found, the whole loop continues till the end of the range.

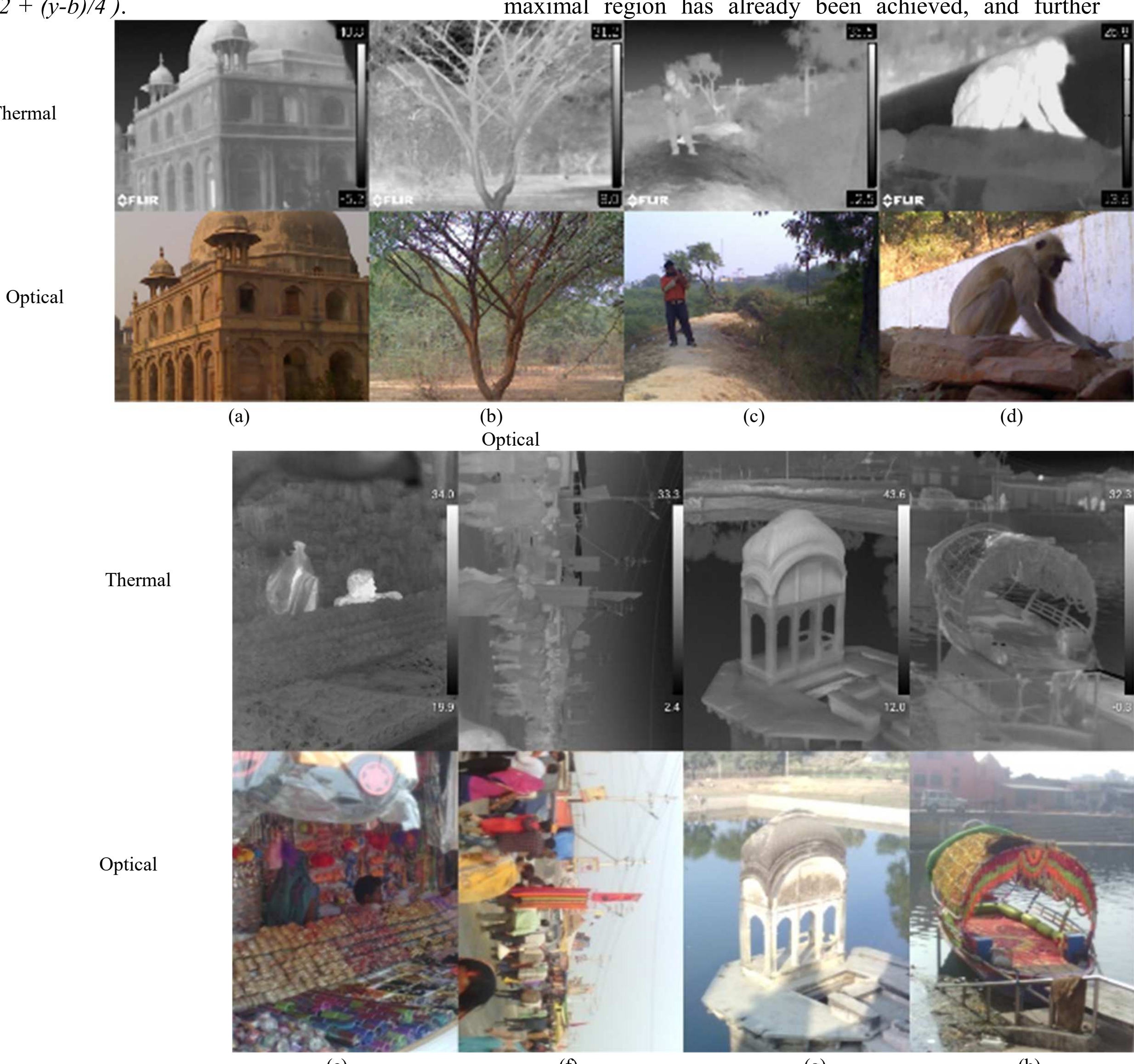

Fig. 3. Registration results for different classes of data. Figures (a)-(d) are images obtained via Flir E40 thermal imager and figures (e)-(h) are images obtained via Sonel KT400 thermal imager

Thus, we introduced a secondary check to this wherein the checking process only continues until 3 full rows if the checking range is skipped without any update to the maximal region. We base this on the assumption that the check for the maximal MI region happens gradually, wherein the optimal region is found out by closing the gap in maximal region. Thus, if there is a break in the check region, it means that the maximal region has already been achieved, and further comparisons need not be performed. This brings in a very major change in the time complexity, reducing the average time by a factor of 91%. We name this *Registration Algorithm 2*.

However, this could be reduced further if we take a region that is 30 pixels away from the edges of the thermal image in each direction. This is because we can see that the thermal band is present in all thermal images towards the right most region. Not only that, we try to capture the images with maximal information regions at the centre of the thermal image. Thus, the trimming does not result in a significant loss of information contained in the region to be compared, providing the same output as the original method. Finally, we

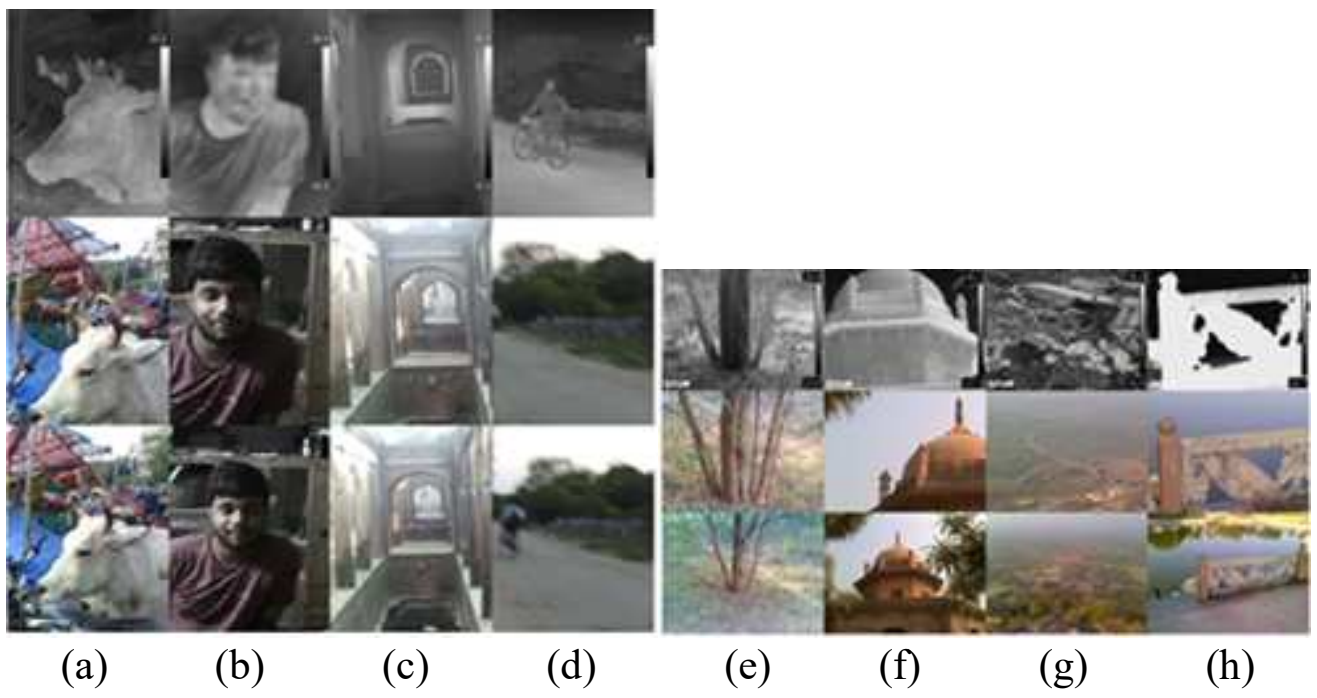

(a) (b) (c) (d) (e) (f) (g) (h)

Fig. 4. Images where the Registration algorithm is unable to provide a perfect match of thermal-optical pairs in case of images captured via the Sonel thermal imager for (a) – (d) and those captured by the FLIR thermal imager for (e) - (h)

cut out the registered optical image from the rescaled RGB matrix as:

$$Image_i = (x_i - 30:(x_i - 30) + a, y_i - 30:(y_i - 30) + b) \quad (4)$$

This process brings down the registration time complexity further by a factor of 32% from the previous iteration. We call this *Registration Algorithm 3*, and is the final registration algorithm we are proposing in this work. We include a bar graph showing the time complexity comparison of *Registration Algorithm 1-3* in Fig. 2. We include the complete *Registration Algorithm 3* in the Supplementary Materials Section. In the Supplementary Materials *Registration Algorithm 3*, the shape of the each optical image is *(length_optical, height_optical)* and that of each thermal image is *(length_thermal, height_thermal)*. *MI* refers to Mutual Information.

It needs to be understood however, that our method does not work for all pairs of thermal-optical images. This is covered in more detail in Section 3.

## III. RESULTS AND DISCUSSION

We provide a few results of our registration method in Fig. 3, where we try to include data from all 5 classes we have in our database.

It can be observed from Fig. 3 (f) that when we photograph crowds, due to the constantly moving foreground, composed of moving humans, the image can still be registered if the background is sufficiently distant from the imager, resulting in a stationary frame of reference. This is also the case for Fig. 3 (d), where there is a slight change in the photograph due to the change in perspective caused by the different locations of the optical and the thermal imager. This can be noticed if we see the slight section of the wall above the animal in the photograph. However, due to the animal itself occupying a majority of the photograph, the images still get registered.

While our method works with most data, it fails in the case of data where the subject is moving, the thermal and the optical sensors are unable to capture the same image. This can be seen in Fig. 4 (b) and (d). This occurs because there is a lag in the operation time of the thermal and the optical sensors

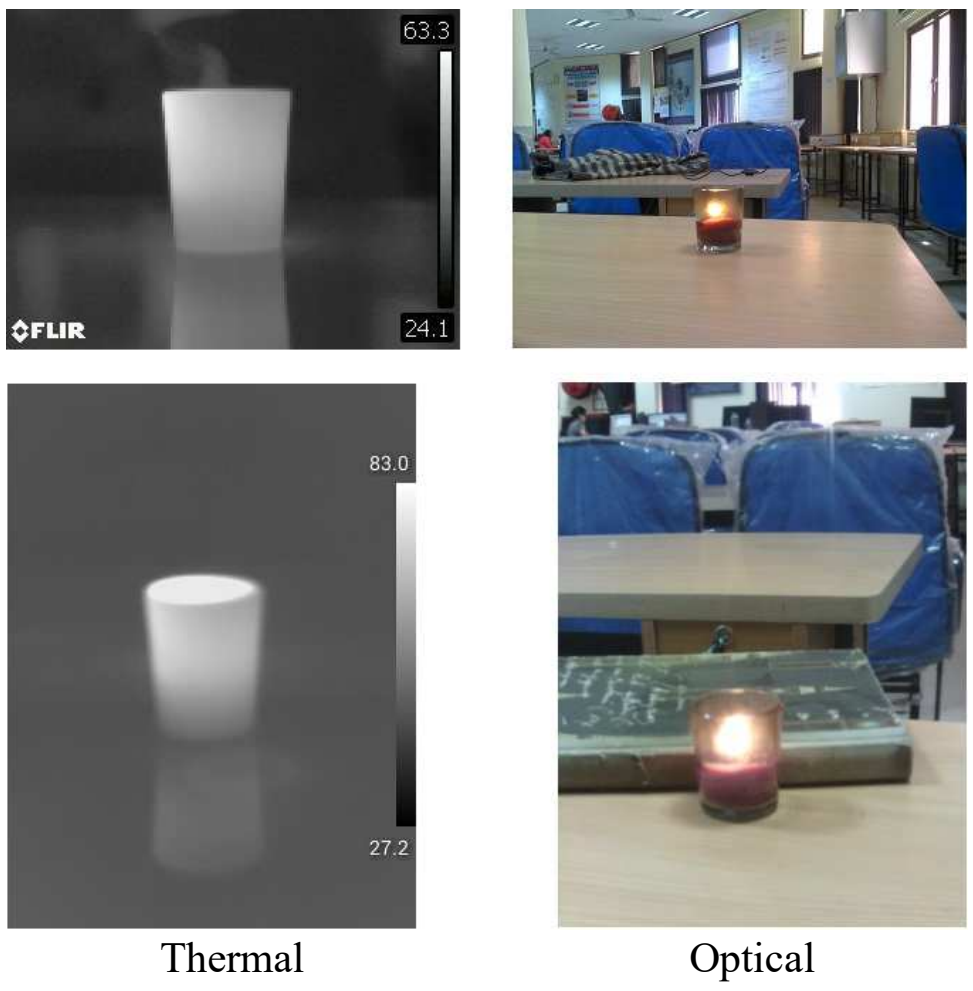

Thermal Optical

Fig. 5. A comparison of the off-centre parallax error for objects at photographed at around 40 cm distance from the thermal imager (a point where both Sonel and FLIR thermal imagers have a clear picture of the object being captured), focused manually via the included focusing ring present in the imagers. We measured the pixel positions of 2 corresponding pixels in the thermal and the resized optical images and they are (113, 277) for the Sonel thermal imager and (294, 292) for the FLIR thermal imager in the resized optical images; both being well within the starting value of (16, 18) and (105, 73) respectively for the Mutual Information based comparison being made for the registration method.

as the 2 work serially. Subsequently, the data captured is different in the 2 cases.

The second case where the registration module fails to provide an accurate registration for an input is in case of a parallax error, as can be observed in Fig. 4 (c). The optical and the thermal sensors are physically located at different locations in both the thermal imagers we have used for creating our database. This creates a shift in the image being captured. We observed that while this does not create much of an error when the objects being captured are far away, at nearby distances, it becomes a major issue. In fact, the image captured becomes completely different when the object is very close to the imager, creating a shift in perspective, thus making the algorithm we have provided being unable to provide a good match. We can notice that in Fig. 4 (c), the arches change in their perspective between the thermal and the optical images altogether and in Fig. 4 (f), we can see that the leaves on the right side of the structure change the image between the thermal and the captured optical images.

Another problem that we have faced is that some images are highly bleached. By 'bleached', we mean that there is not sufficient difference in the different surfaces of the object in the thermal image. This is primarily the case when we deal with objects which are left for too long at a certain temperature (for example, an animal sitting in the sun). This can be seen in Fig. 4 (h) and (a). In the first one, we can see that due to the extreme difference between the foreground and the background, there is a distinctive loss in features in the image, while in the second case, the background becomes blurry, creating a case where the Mutual Information based algorithm is unable to get enough unique information to create a good match.

Finally, the last case where we see problems with registration is where there is a loss of data due to the sensor being unable to capture precise data. This can be seen in Fig. 4 (e) and (h). In case of Fig. 4 (e), we can see that the small branches in front of the tree are not captured in the thermal image, thus creating a different image for the algorithm, while in (h), the optical sensor is unable to capture the correct image corresponding to a thermal map due to fog and imperfect illumination.

All of our data is available for public use in IEEE Dataport at [1]. We have also included the raw data for the unregistered images in the database so that others may yet use them in further researches in other avenues.

## IV. CONCLUSION

We have presented a method which has been tested with data from Prayagraj (Allahabad) city in India for the modern setting based images and jointly from Chitrakoot and Prayagraj for the historical buildings images and the greenery data. The crowd data is collected from the Maha Kumbh Mela 2019 that occurs once every 12 years at Prayagraj. It is the biggest fair on earth and to our knowledge, this is the only dataset that contains images from it. The data was collected over a period of 2 years and collated together.

The registration method we are proposing is not only computationally lightweight, it is robust enough to work across 2 different thermal imagers. These have widely differing image resolutions and are used to capture images from varying scenarios, which show that our method can work irrespective of the domain of the image captured.

All our registered images are available in our public dataset, published at [1].

## ACKNOWLEDGMENT

We would like to thank the DIPR Lab, DRDO, Government of India for funding this research under grant number 2535/DIPR-II/MD/CARS-01 and Computer Visions and Biometric Lab (CVBL), IIIT Allahabad for providing the necessary facilities to conduct it. We would also like to thank Prof. Gaurav Sharma, University of Rochester, New York, USA for his expert guidance on the research work and Mr. Nand Kumar Yadav, IIIT Allahabad, without whose help the data collection drive would never have been completed.